# Fully Integrated Perovskite Polaritonic Circuits with Tunable Lasing and Nonlinear Amplification

Antonio Fieramosca[1,*], Vincenzo Ardizzone[1], Rosanna Mastria[1], Laura Polimeno[1], Lorenzo Dominici[1], Umberto Dellasette[2], Andrea Notargiacomo[3], Marialilia Pea[3], Raffaella Polito[3], Eugenio Maggiolini[4], Rughianah Gohar Ashfaq[1], Fabio Bruni[5], Elisa Fardelli[5], Simona Sennato[6], Francesco Todisco[1], Dario Gerace[7,1], Dario Ballarini[1], Milena De Giorgi[1], Ilenia Viola[2,*] and Daniele Sanvitto[1,8,*]

**Affiliations**

[1] CNR NANOTEC Institute of Nanotechnology, Via Monteroni, Lecce 73100, Italy

[2] CNR NANOTEC Institute of Nanotechnology, Sapienza Università, P.le A. Moro 2, Rome I-00185, Italy

[3] CNR IFN Institute for Photonics and Nanotechnologies, Via del Fosso del Cavaliere, 100, 00133 Roma, Italy

[4] Institute of Semiconductor and Solid State Physics, Johannes Kepler University, Altenbergerstraße 69, Linz, Austria

[5] Dipartimento di Scienze, Università degli Studi Roma Tre, Via della Vasca Navale, 84, 00146 Rome, Italy

[6] CNR ISC Institute for Complex System, c/o Physics Dept, Sapienza Università, P.le A. Moro 2, Roma I-00185, Italy

[7] Dipartimento di Fisica "A. Volta", Università di Pavia, via Bassi 6, 27100 Pavia, Italy

[8] Visiting professor at Division of Physics and Applied Physics, School of Physical and Mathematical Sciences, Nanyang Technological University, Singapore 637371, Singapore

*To whom correspondence should be addressed. Emails: antonio.fieramosca@cnr.it, ilenia.viola@cnr.it, daniele.sanvitto@cnr.it

# Abstract

Photonic integrated circuits are emerging as a key technology for compact and energy-efficient optical information processing. Yet, their practical implementation remains limited by the intrinsically weak optical nonlinearities of conventional materials, which demand high power and large footprints to achieve significant nonlinear responses. Exciton-polaritons, hybrid light-matter excitations of semiconducting materials, offer a promising solution by combining strong optical nonlinearities with the high speed and large scalability typical of photonic devices. However, despite their potential, working on-chip polaritonic elements demonstrating room temperature coherent lasing, controllable nonlinear propagation, or amplification have remained elusive. Here we demonstrate a fully integrated perovskite polaritonic circuit that overcomes these limitations. Using a single-step microfluidic lithographic technique, we realize waveguide circuits with integrated gratings that simultaneously act as couplers and mirrors, forming in-plane Fabry-Pérot cavities. These structures support robust in-plane polariton lasing between gratings, yielding coherent emission along the waveguide. Furthermore, we observe clear signatures of strong nonlinear self-phase modulation and, for the first time, optical amplification of guided polaritons at room temperature. Our simple, scalable platform opens the way to low-power, highly nonlinear optical circuits for integrated photonics and neuromorphic architectures operating at room temperature.

# Introduction

Integrated photonic devices have emerged as a powerful platform for the next-generation information technologies, and the development of efficient and scalable integrated photonic circuits has become a central focus of state-of-the-art research. Photonics can achieve ultrafast signal propagation with low transmission losses, broadband modulation, and highly efficient information processing, even outperforming conventional electronic devices [1–3]. In this context, it is crucial that the same architecture may support a variety of integrated components, including on-chip coherent light sources, frequency-comb generators, modulators, amplifiers, and switches, in order to achieve the necessary dynamical reconfigurability. However, the effectiveness of all-optical devices depends critically on the ability to harness optical nonlinearities. Since conventional platforms based on linear dielectrics materials or lithium niobate ($LiNbO_3$) exhibit intrinsically weak nonlinear responses [4–6], structures supporting exciton-polaritons have been proposed as an effective alternative. These hybrid light-matter excitations, formed under non-perturbative radiation coupling between excitons and purely photonic eigenmodes, combine photon-like propagation with a much stronger nonlinear response, hence satisfying the essential requirements for ultrafast all-optical information processing with minimal thermal load [7–11].

Although exciton-polaritons have been extensively studied in vertically confined microcavities [12–14], planar architectures such as polariton waveguides [15–18] have recently attracted increasing interest owing to their ability to confine light within smaller mode volumes, thereby enhancing exciton-photon coupling, increasing polariton group velocities, and facilitating on-chip integration. When combined with suitable active materials exhibiting large exciton binding energies, such as organic, perovskites, transition metal dichalcogenides, and wide-bandgap semiconductors like GaN and ZnO, polariton operation can be sustained at room temperature [9,19–25]. Perovskites, in particular, stand out for their exceptional and tunable optoelectronic properties and ease of processing, which eliminates the demanding fabrication procedures typically required for conventional semiconductors. In addition, their strong polariton nonlinearities, comparable to that of GaAs-based systems, makes them an increasingly attractive platform for polariton-based applications at room temperature [26–29]. However, the fabrication of photonics components in perovskite materials, notably waveguides, grating couplers, interferometers, and beam splitters, demands precise control on the crystal morphology, since well-defined edges and smooth surfaces are necessary to achieve satisfactory performance [30–32]. Nevertheless, the realization of such structures is highly demanding, since it typically involves post-processing steps employing polar solvents that can deteriorate both the structural integrity and the optical quality of the perovskite. Despite recent progress, efficient in- and out-coupling of light remains a major bottleneck. The realization of coherent on-chip tunable light sources as well as optical amplification of propagating polariton signals, an essential function for active circuitry, still remain challenging goals and very much unexplored. Overcoming these limitations is crucial to achieving robust, room-temperature, scalable, and fully integrated nonlinear photonic devices.

Here, we address these critical challenges by developing a fully integrated perovskite-based polaritonic circuit in which waveguides and grating couplers are simultaneously defined within the active materials through single soft-lithography step directly from the liquid phase, eliminating the need for any subsequent post-processing. Using a surface-tension-assisted microfluidic growth process [33–35], we transfer the geometries defined by a silicon master onto $CsPbBr_3$ perovskite crystals via an elastomeric replica, thereby producing hundreds-of-microns-long waveguides with micrometric widths and integrated subwavelength gratings with tunable periodicity and filling factors. The resulting structures exhibit strong nonlinear optical effects, including self-phase modulation of resonantly injected polariton wavepackets. Under off-resonant pumping, we observe polariton lasing enabled by longitudinal optical feedback provided by the integrated gratings, which act as partially reflecting mirrors forming an in-plane Fabry-Pérot cavity. In-plane laser emission propagates efficiently along the waveguide, and, most importantly, we demonstrate that the lasing wavelength can be readily tuned by simply modifying the coupler geometry. Finally, we also demonstrate efficient optical amplification in a two-beam experiment, in which a resonantly injected propagating polariton wavepacket is amplified through stimulation from the exciton reservoir created by an additional off-resonant beam. Remarkably, the observed threshold-like amplification and nonlinear response resemble key features of spiking neurons, offering a promising route toward the development of neuromorphic polariton devices capable of all-optical computation and signal processing [36,37].

Our results represent the first demonstration of room-temperature polaritonic circuits capable of generating tunable coherent emission, efficient amplification, and precise control over coupling properties, paving the way for groundbreaking advances in integrated polariton devices.

**Results**

CsPbBr$_3$ perovskite waveguides with various geometries were directly fabricated on glass substrates using a modified soft-lithography approach that combines surface tension-driven lithography with microfluidic processing from a liquid precursor solution (Figure 1a) [38–42]. This hybrid process enables the direct formation of hierarchical architectures through controlled self-assembly under saturation conditions, yielding micrometer-scale waveguides decorated with integrated nanometric grating arrays. Elastomeric polydimethylsiloxane (PDMS) molds were produced by replica molding from a silicon master patterned by electron-beam lithography (EBL, see Methods). The masters feature a multi-level, positive waveguide design that incorporated microwires and interferometer-like structures with lateral dimensions between 2 and 30 µm and a height of 250 nm (Figure 1b and Figure S1). Nanograting matrices were also etched on the upper surface of these structures, covering areas of about 10 × 10 µm$^2$ with variable pitch, filling factor, and spatial position.

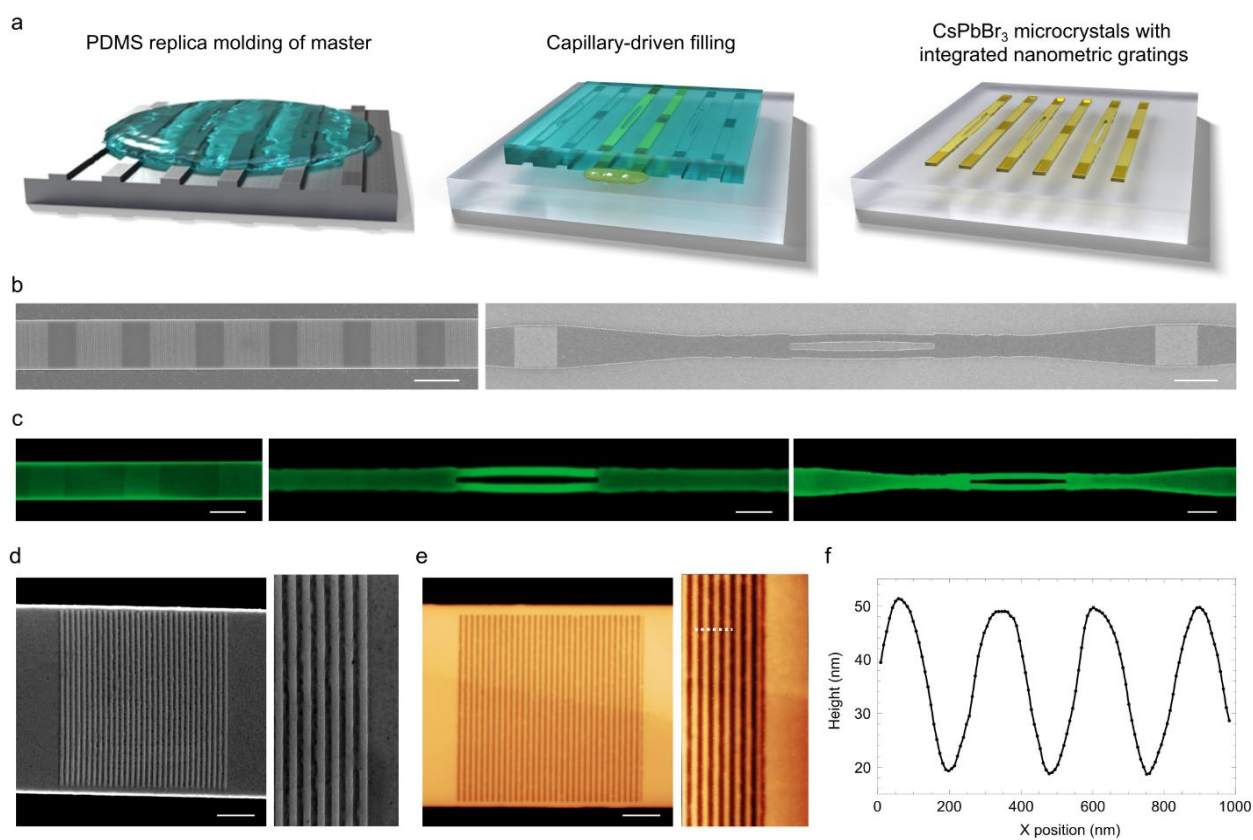

**Figure 1**: **a)** Schematic illustration of the fabrication process of CsPbBr$_3$ waveguides via microfluidic surface-tension-driven lithography. The process follows three main steps: (1) replica molding in PDMS of the waveguide master in PDMS (left); (2) capillary-driven distribution of the precursor solution along the microstructured channels under saturated conditions and controlled temperature (center); and (3) formation of CsPbBr$_3$ microcrystals with the desired morphology and integrated nanometric optical gratings (right); **b)** SEM images of representative waveguides with integrated optical gratings fabricated on the silicon master via EBL; **c)** Confocal images of CsPbBr$_3$ perovskite waveguides with integrated gratings produced by the modified single-step soft-lithography technique described in (a); **d), e)** SEM and AFM images displaying a complete grating coupler directly replicated onto the CsPbBr$_3$ perovskite structure. Right insets display high-resolution views of the gratings. **f)**, AFM cross-section profile of the realized grating couplers, corresponding to the white dashed line shown in (e, right inset). Scale bars: (b, c) 10 µm; (d, e) 2 µm.

The precursor solution was deposited at the entrance of the PDMS pattern placed in conformal contact with a glass substrate. Capillary forces and surface interactions guided the solution through the microfluidic channels. Within the confined geometry, the small precursor volumes and the gradual increase in concentration and viscosity gradients promoted perovskite nucleation through surface-tension-driven self-assembly at the elastomer-air interface. Crystallization proceeded under controlled temperature and vapor conditions. After PDMS removal, high-resolution and reproducible $CsPbBr_3$ crystals were obtained directly patterned on glass (Figure 1c). The controlled growth environment preserved the master's structural features and ensured high crystalline quality (Figure 1d–f), enabling the successful replication of diverse grating geometries with varying filling factors (Figure S2) and pitches (Figure S3) in the perovskite layer. This single-step process defines both waveguide channels and grating couplers simultaneously, achieving aspect ratios (height/width) from 0.01 to 0.4 with excellent structural precision [33]. The low adhesion of PDMS minimizes edge roughness and defect formation, enabling accurate control over crystal size and the realization of clean interfaces. Overall, this hybrid soft-lithography route overcomes key limitations of conventional fabrication, eliminating multi-step lithography, post-processing damages, and refractive-index mismatches, while drastically reducing fabrication cost and time. Moreover, the same silicon master can be reused for multiple growth cycles, paving the way toward scalable, high-quality integrated perovskite photonic devices.

To investigate the optical properties of the fabricated structures, we performed photoluminescence (PL) measurements under pulsed femtosecond excitation (200 fs, 480 nm, 10 kHz, see Methods). The waveguide was excited between two gratings, and momentum-resolved PL measurements revealed the presence of propagating and counterpropagating waveguide modes in both TE and TM polarizations, as shown in Figure S4 (see Methods). The resulting energy versus in-plane momentum PL map collected by spatially selecting the emission from only one of the nearest outcouplers is shown in Figure 2a, while the typical full real-space PL emission is shown in the inset. All the results discussed in the text refer to the TE polarization. Two strongly coupled modes are clearly visible (green dashed lines), which we attribute to the fundamental ($TE_0$) and first-order ($TE_1$) lower polariton (LP) waveguide modes. The presence of these two modes is consistent with a waveguide thickness of 250 nm and a grating depth of approximately 50 nm, as confirmed by simulations performed by using rigorous coupled-wave analysis (RCWA), and presented in Figure S5 for different grating pitches. Although the upper polariton branch is not visible in perovskite-based strongly coupled systems (as previously reported [43–45]), the characteristic dispersion of the waveguide modes, which should be linear in the weak coupling regime (white dashed lines), shows an increasing bending as the energy approaches the exciton resonance ($E_{exc}$ = 2.4 eV, red solid line), providing clear evidence of the strong radiation-matter coupling and the formation of lower polariton branches. The experimental data were fitted by using a two-coupled-oscillator model, yielding Rabi splitting energies of $\Omega_1$ = 250 meV and $\Omega_0$ = 215 meV for the $TE_1$ and $TE_0$ modes, respectively. It is important to note that the $TE_1$ mode appears brighter and exhibits a broader linewidth in momentum space as compared to the $TE_0$ mode. This can be understood by considering the electric field distributions shown in Figure S5, where the corresponding electric field amplitudes of the first two TE-polarized modes are presented. The $TE_1$ mode has one of its field maxima located closer to the top surface of the waveguide, where the grating coupler is patterned, which results in more efficient coupling to free space and a broader dispersion.

To assess the presence of a nonlinear optical response we performed a resonant pulsed excitation experiment. The pumping laser ($E_{inj}$ = 2.3 eV) was tuned to the brightest mode, i.e., the $TE_1$, with a beam waist in momentum space smaller than the separation between the two modes, thereby ensuring selective excitation of this mode along a specific in-plane direction. The signal outcoupled from the first adjacent grating was then monitored in momentum space as reported in Figure 2b. The corresponding real-space propagation map and the momentum-space dispersions for three different excitation energies are presented in Figure S6, which clearly demonstrates that the resonant laser is efficiently coupled into and extracted from the $TE_1$ polariton mode. Under these experimental conditions, we monitored the spectral modifications of the propagating beam as a function of the input fluence, for a fixed excitonic fraction (X = 0.65, Figure S7) and a propagation distance of approximately 25 μm (distance between the in- and out-coupler). The collected spectra, shown in Figure 2c for two different pumping fluences, reveal that at low pumping fluence the spectrum remains essentially unmodified during propagation (bottom panel). In contrast, as the pumping fluence increases, a pronounced spectral broadening appears (top panel). The complete energy vs pumping fluence map is presented in Figure 2d, showing the continuous spectral broadening and the emergence of new frequency components. The observed spectral evolution can be interpreted as the combined action of self-phase modulation, nonlinear group velocity dispersion, and self-steepening, as recently discussed in both $MAPbBr_3$ [46] and GaAs-based polariton waveguides [47,48]. Although an in-depth investigation of these nonlinear effects lies beyond the scope of this work, our results are fully consistent with those already observed in $MAPbBr_3$ polariton waveguides.

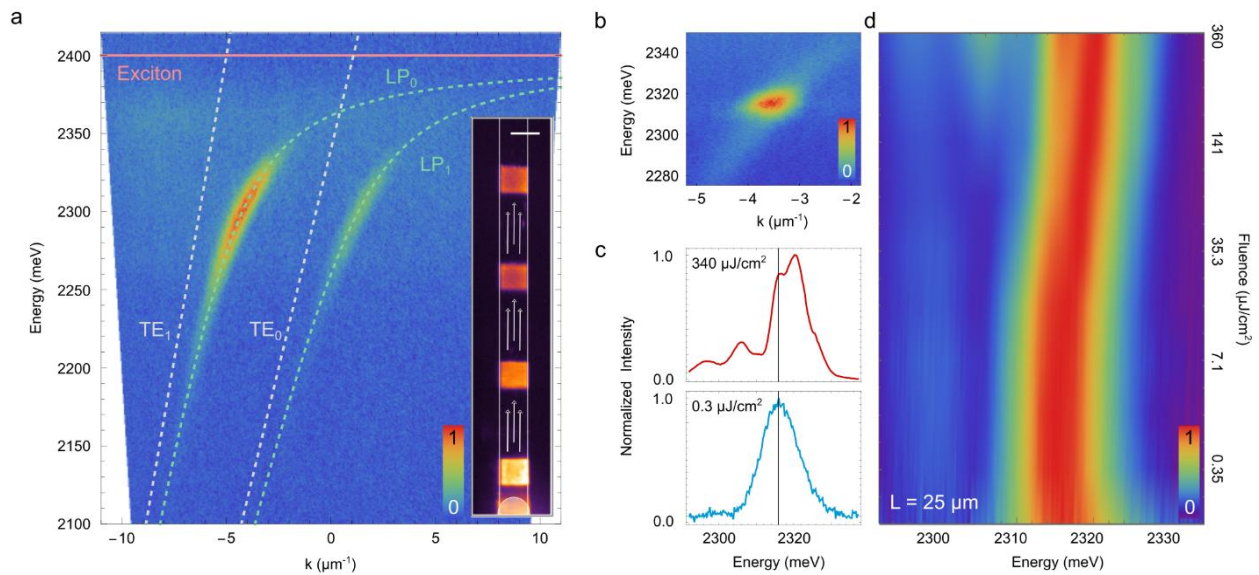

**Figure 2**: **a)** Polarization resolved (TE) momentum-space PL map from one of the waveguide gratings. Two polariton modes are visible (green dashed lines) and marked $LP_0$ and $LP_1$ respectively. The two polariton modes arise from the strong coupling between two photonic modes ($TE_0$ and $TE_1$, white dashed lines) and the excitonic resonance (horizontal solid line). Inset: real space PL image of the waveguide in which the emission from several adjacent gratings is visible; the white circle in the bottom part indicates the pumping spot; **b)** Momentum-space map showing the resonant signal outcoupled from the first adjacent grating. Note that a weak off-resonant beam is added to simultaneously visualize the polariton dispersion; **c)** Nonlinear propagation of a resonantly injected pulse; The two graphs show the spectra of the transmitted pulse for low (bottom) and high (top) pumping fluence, respectively (see labels). When the pump fluence is increased, self-phase modulation takes place. **d)** Bidimensional map of the self-phase modulation effect, in which each row represents the spectrum of the transmitted pulse, pump fluence being reported on the vertical axis.

Notably, signatures of self-phase modulation in MAPbBr$_3$ waveguides are only present for lengths above 50 µm, while in our waveguides we can observe the effect at lower fluences (130 µJ/cm$^2$) and shorter lengths (25 µm), confirming the superior nonlinear performance of our CsPbBr$_3$ platform.

Next, we show that our perovskite waveguide architecture can serve as a highly efficient and spectrally tunable on-chip laser source. As previously demonstrated, in the absence of external optical resonators, lasing in perovskite microstructures can arise from the optical feedback provided by sharp crystal facets acting as Fabry-Pérot cavities [31,32,49,50]. In the case of a waveguide, this feedback occurs predominantly in the transverse direction [31], and because it develops perpendicular to the waveguide axis, the resulting signal is not efficiently coupled into the guided modes, thereby hindering effective in-plane propagation. Moreover, this lasing emission lacks spectral tunability, as it is governed by the intrinsic balance between optical gain and loss within the material. Alternative strategies have involved isolated perovskite blocks evanescently coupled to active waveguide channels [32], but suffer from limited efficiency due to the weak overlap of evanescent fields. Consequently, these approaches neither provide precise spectral control nor efficient coupling of coherent emission along the propagation direction, fundamentally limiting their scalability for integrated photonic applications. In contrast, the grating couplers adopted in our work not only enable efficient detection of TIR-confined modes but also act as partially reflecting elements providing longitudinal optical feedback. A similar waveguide-based polariton laser was demonstrated in a GaAs-based waveguide at cryogenic temperature [18] and in GaN-based waveguide at room temperature [51], although in this case the reflective elements were fully etched. Importantly, our approach achieves this by employing a very simple and direct growth process that produces waveguides with integrated couplers already in the desired shape, in contrast to the more complex fabrication required for GaAs or GaN platforms.

In order to verify the formation of such a longitudinal optical feedback, the perovskite waveguide was off-resonantly pumped between two couplers. The real-space PL map is shown in Figure 3a for low (left, P < P$_{th}$) and high (right, P > P$_{th}$) pumping power, respectively. At low pumping power, the two outcouplers are barely visible without spatially excluding the area under the excitation spot, as only a relatively small portion of the emitted light is effectively coupled into the waveguide modes under off-resonance pumping. In contrast, at high excitation power, a pronounced signal emerges from both outcouplers, immediately suggesting that a longitudinal feedback is taking place. The corresponding momentum-resolved maps are reported in Figure S8. To further investigate the onset and evolution of the lasing emission, we performed a real-space energy-resolved measurement and we evaluated the emitted intensity collected from one of the two outcouplers for both the lasing signal (blue symbols, integrated between 2.31-2.32 eV) and the PL (red symbols, integrated between 2.3-2.31 eV), as shown in Figure 3b. As it is evident, PL intensity increases linearly with pump fluence, whereas the lasing signal exhibits a clear threshold at approximately P$_{th}$ = 50 µJ/cm$^2$, consistent with previous reports on similar perovskite materials [27,49,52,53]. The presence of the photonic stop band associated with the gratings can be directly observed in the inset of Figure 3b, which displays the momentum-integrated emission profiles below and above the lasing threshold. Below threshold, the emission intensity collected through the mode would normally follow the PL spectral profile, showing a monotonic decrease in the considered energy range. However, the extracted profile exhibits a distinct dip (bottom part of the inset), and above threshold, a sharp and narrow lasing peak emerges within this spectral region (top part of the inset), corresponding to the stop band of the longitudinal

Fabry-Pérot cavity [18]. Moreover, it is important to note that the lasing mode blueshifts with increasing power, a characteristic signature of polariton lasing, and that an additional Fabry-Pérot mode appears, still located inside the stop band, as can be seen in Figure S9. The emergence of multiple modes is consistent with the relatively long cavity formed between the two couplers, which supports different longitudinal resonances capable of achieving population inversion.

In order to demonstrate the spectral tunability of the lasing emission, we have tuned the periodicity within the integrated couplers by keeping constant their mutual distance. This modification directly controls the longitudinal cavity modes and, consequently, the lasing energy.

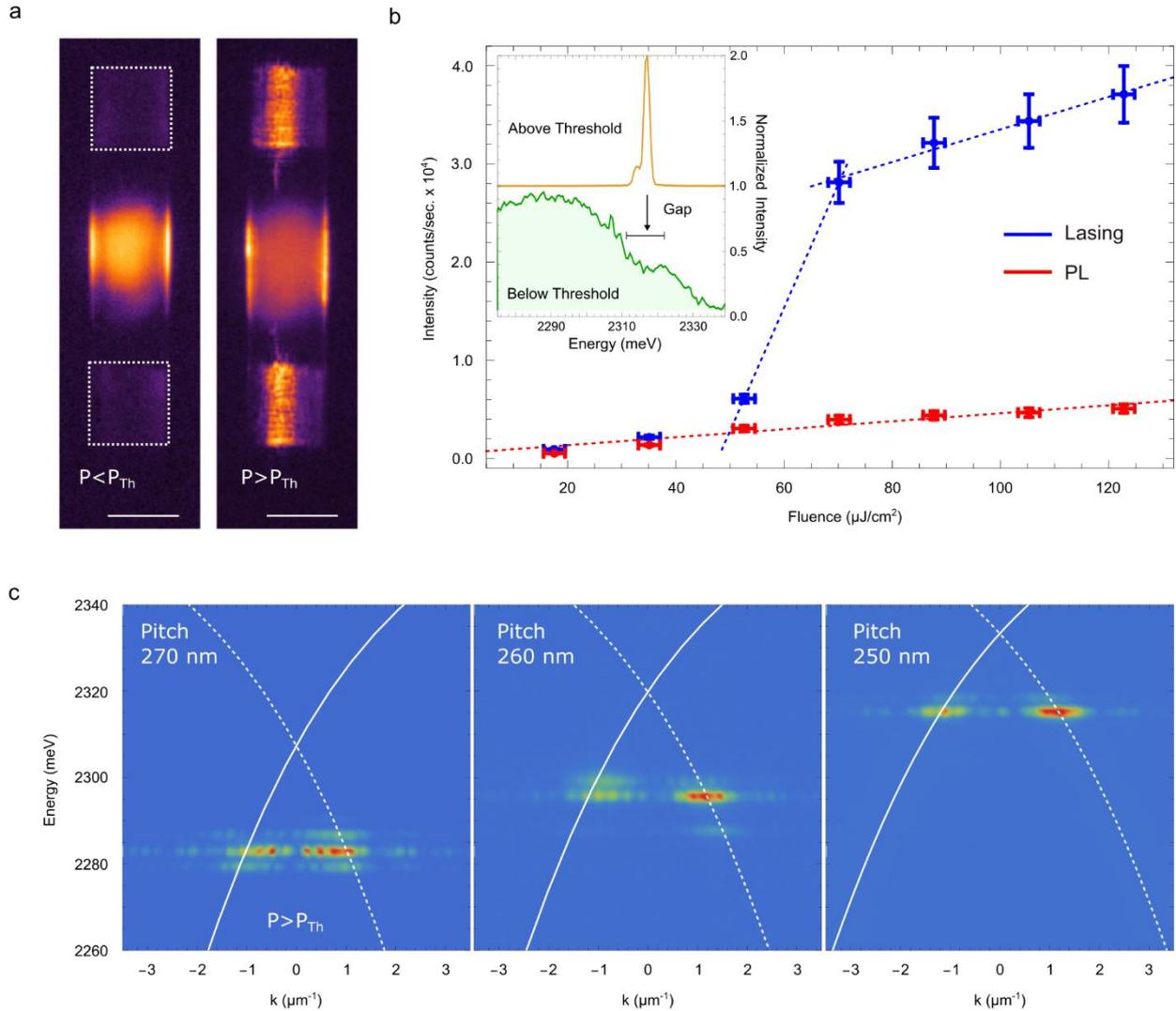

**Figure 3**: **a)** Real space image of a Fabry-Pérot in-plane cavity formed by two gratings (highlighted by the dashed squares); below threshold (left panel) the emission is dominated by exciton recombination below the pump spot while above threshold (right panel) the lasing mechanism takes place and the emission from the two gratings is stronger than the PL not coupled to the guided mode; **b)** Emitted intensity integrated in real space energy-resolved measurements for both the lasing signal (blue symbols, integrated between 2.31-2.32 eV) and the PL (red symbols, integrated between 2.3-2.31 eV), as a function of pumping fluence. The inset shows the photonic stop band associated with the two gratings: below threshold, a distinct dip is visible (bottom part) in momentum-integrated emission profiles, while above threshold, the lasing peak emerges within this spectral region. **c)** Above threshold lasing emission from three different cavities formed by couples of gratings having different pitches (respectively 270, 260 and 250 nm); when the pitch is modified the energy of the photonic stopband changes and the lasing energy shifts accordingly.

As shown in Figure 3c, three different grating geometries with pitches of 270, 260, and 250 nm result in a progressive blueshift of the lasing mode from approximately 2.28 eV to 2.315 eV. We stress that the lasing signal collected from the first outcoupler next to the grating-pair cavity element remains almost comparable in intensity to that emitted from the cavity itself, as shown in Figure S10, thus providing clear evidence of efficient and long-range propagation of the lasing mode along the waveguide. These results demonstrate that the lasing energy can be precisely tuned by properly tailoring the grating design, thereby shifting the energy position of the narrow photonic stop band. The presence of two such gratings gives rise to a cavity effect along the waveguide axis, creating an energy stop band in the guided-mode dispersion. Within this band, a series of Fabry-Pérot resonances emerges through multiple reflections between the gratings, thus forming a resonator that confines and recycles photons. This confinement enables stimulated emission on a well-defined polariton mode, and it provides a straightforward route to achieve specific lasing energies without additional processing or material modification.

In the following section we demonstrate another key functionality of our system: the optical amplification, which has not been previously achieved in similar perovskite polaritonic platforms working at room temperature. This capability is essential for integrated photonic circuits as it compensates for propagation and coupling losses, thus enabling efficient on-chip light transport. For this specific experiment we employed a modified interferometer-like waveguide in which the input and output gratings are positioned approximately 80 μm apart. The relatively long separation was deliberately designed to suppress the longitudinal optical feedback previously observed, allowing us to focus exclusively on the amplification process. At the same time, this geometry demonstrates the ability of our fabrication technique to realize complex on-chip structures. Indeed, before the Y-shaped splitting region, the waveguide incorporates a sequence of lateral modulations that efficiently project the guided mode into both branches. Achieving such precise shaping requires careful design and fabrication control, underscoring the versatility of our approach for realizing advanced integrated photonic geometries.

In order to investigate optical amplification, we resonantly injected a pulsed beam (hereinafter referred to as the signal, $E_{signal}$ = 2.3 eV) through the input grating and monitored the transmitted signal at the output. A second pulsed beam (hereinafter referred to as the amplifier, $E_{amplifier}$ = 2.58 eV) was off-resonantly exciting the center of the waveguide, far from the input grating, where it pumps both arms of the interferometer to generate an incoherent exciton population, as schematically shown in Figure 4a. The output intensity was recorded by scanning the time delay between the two pulses, keeping the signal with a constant intensity and changing the fluence of the amplifier. Figure 4b shows the collected intensity as a function of time delay for three different amplifier fluences, 5, 18 and 31 μJ/cm$^2$, indicated by the blue, orange and green curves, respectively. The residual PL signal generated by the amplifier was collected separately and subtracted from the measurements where both beams are simultaneously present (see also Methods). As it can be seen, at low pump fluences, the signal intensity at the output grating remains constant, indicating that the amplifier does not affect the transmission of the signal throughout the structure. However, as the amplifier fluence increases, the transmitted signal exhibits a pronounced enhancement when the two beams are temporally synchronized (Δt = 0 ps), which then gradually decays and vanishes within approximately 50 ps. The spectral profiles of the amplifier (red), signal (green), and amplified-signal (blue) collected at Δt = 0 ps are shown in Figure 4c for three different amplifier fluences. These spectra clearly demonstrate the onset of amplification once a sufficient particle density is reached. To further elucidate the effect, power-dependent measurements at Δt = 0 ps were performed, as shown in Figure 4d. Here, the red dots (amplifier) represent the PL intensity generated by the off-resonant amplifier alone, which increases as a parabolic-like function of the pump fluence (Figure S11), whereas the blue points (amplified-signal) correspond to the intensity measured in the presence of both the signal and the amplifier beams, showing clear amplification with a marked threshold-like behaviour. The observed trend not only confirms the occurrence of optical amplification, but it also points towards an amplification process consistent with stimulated emission, where the presence of the probe stimulates particles from the incoherent exciton reservoir created by the amplifier. This process may be assisted by acoustic phonons, which have recently been identified as playing a fundamental role in perovskites [54]. We then evaluated the optical gain as shown in the inset of Figure 4d. The gain is defined as the ratio of the amplified-signal intensity (amplified-signal), corrected by subtracting the amplifier intensity (amplifier), to the signal alone intensity (signal), expressed as: $Gain = \frac{(Amplified\ Signal - Amplifier)}{Signal}$. As it can be seen, the gain reaches values greater than 7. To the best of our knowledge, these results represent the first experimental demonstration of optical amplification in a perovskite waveguide system operating at room temperature.

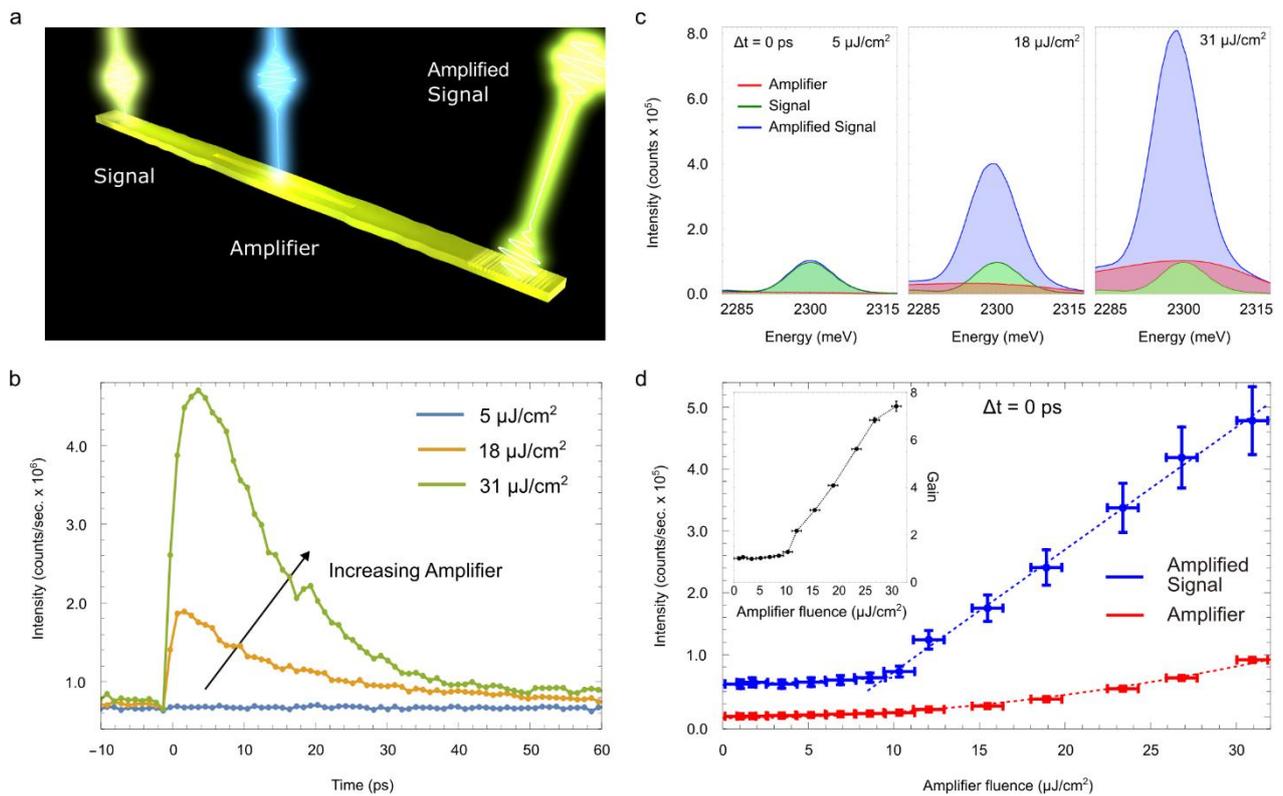

**Figure 4**: **a)** Sketch of the integrated optical circuit used to measure polariton amplification: the off-resonant beam (Amplifier) creates an exciton reservoir while the resonant beam (Signal) excites a polariton state propagating along the waveguide. At the output grating the signal is amplified (Amplified Signal) thanks to stimulation from the exciton reservoir; **b)** Collected intensity of the amplified signal as a function of time delay for three different amplifier fluences, 5, 18 and 31 μJ/cm$^2$; **c)** Spectral profiles of the amplifier (red), signal (green), and amplified-signal (blue) collected at Δt = 0 ps for three different amplifier fluences, 5 (left panel), 18 (middle panel) and 31 μJ/cm$^2$ (right panel), respectively. **d)** Power-dependent measurements at time delay 0 ps; red dots represent the intensity of the amplifier alone, while blue dots the intensity of the amplified signal as a function of the amplifier fluence. The inset shows the optical gain as defined in the main text which reaches values greater than 7.

In comparison with other platforms, such as GaAs- or ZnO-based polariton waveguides [17,55], in which amplification has been reported at cryogenic temperatures, our findings mark an important step forward. Indeed, the developed active polariton platform not only operates at room temperature, but also overcomes many of the technological challenges associated with conventional semiconductor systems, as perovskites can be easily processed and deposited in arbitrary geometries on a wide variety of substrates, thus avoiding the demanding fabrication procedures and refractive-index constraints typical of other kind of heterostructures. It is also important to note that, in the adopted configuration, only a relatively small portion of the waveguide is pumped, while the remaining unpumped section is not actively used. Greater amplification could, in principle, be achieved by increasing the spatial overlap between the pump and probe beams.

**Conclusions**

In summary, we have developed a fully integrated perovskite polariton waveguide platform operating at room temperature that combines strong nonlinear propagation, tunable polariton lasing, and optical amplification within a single device. Crucially, our approach enables single-step microfluidic fabrication, where both the active waveguides and the integrated grating couplers are simultaneously defined during the growth of high-quality $CsPbBr_3$ crystals. This provides a straightforward, cost-effective, and scalable route for realizing self-assembled complex polaritonic architectures without the need for specialized vacuum equipment or epitaxial reactors. The developed architecture supports on-chip polariton lasing with wavelength-selective emission enabled by longitudinal optical feedback provided by the integrated gratings. The coherent emission efficiently propagates along the waveguide axis and energy tunability can be achieved by simply changing the couplers geometry. Therefore, this type of polariton laser therefore represents a compact on-chip coherent light source, ideally suited for integrated photonic applications. Beyond lasing, we demonstrate for the first time in a perovskite polariton waveguide system operating at room temperature, optical amplification of guided polariton signals originating through stimulation from an exciton reservoir. The observed threshold-like amplification and nonlinear response resemble key features of spiking neurons, offering a promising route toward the development of neuromorphic polariton devices capable of all-optical computation and signal processing. By engineering the excitation geometry and temporal sequence of input pulses, such polariton-based systems could, in principle, perform neuron-like operations, including signal integration and threshold-triggered emission. These features, combined with room-temperature operation, material tunability, ease of fabrication and cost effectiveness, highlight the potential for perovskite polariton waveguides as a versatile platform in view of realizing advanced nonlinear devices.

## Methods

### Fabrication of the silicon master

The grating and waveguide nano- and microstructures were fabricated on 25 × 25 mm Si(100) substrates using EBL (RAITH Voyager) followed by reactive ion etching (RIE). The grating pattern was defined using a 130 nm-thick PMMA electron-beam resist layer spin-coated onto the silicon substrate and exposed by EBL at a dose of 580 µC/cm$^2$. The exposed resist was developed in a MIBK:IPA (1:3) solution. A 40 nm-thick chromium film, serving as an etch hard mask, was subsequently deposited by electron-beam evaporation, and the remaining resist was removed in acetone. The pattern was transferred for 50 nm into the silicon substrate using fluorine-based RIE ($CF_4$, $CHF_3$, $O_2$, Ar). The chromium mask was then removed by wet chemical etching. An analogous process was employed for the fabrication of the waveguide structures, with optimized parameters as follows: PMMA resist thickness of 360 nm, EBL exposure dose of 280 µC/cm$^2$, chromium mask thickness of 60 nm, and RIE etching depth of 250 nm. Following chromium mask removal, the silicon master was cleaned in a piranha solution for 15 min to eliminate any residual organic contamination.

### Fabrication of multiscale CsPbBr$_3$ Perovskite Waveguides

Microfluidic growth devices were fabricated by conventional soft lithography and replica molding from the silicon master with predefined multiscale geometries. A mixture of PDMS prepolymer and curing agent (10:1, Sylgard 184, Dow Corning, USA) was cast onto the Si master and cured at 140 °C for 10 min. The polymerized PDMS layer was then carefully detached from the master before the use. The patterned area was opened with a razor blade, and the PDMS replica was brought into conformal contact with a clean glass substrate. A 0.42 M precursor solution was prepared by dissolving 106 mg of CsBr and 183.5 mg of $PbBr_2$ (1:1 molar ratio) in 1.2 ml of DMSO. The mixture was stirred at 80 °C on a hotplate inside a nitrogen-filled glove box for 2 h, yielding a clear solution. Immediately before use, the precursor solution was diluted to concentrations in the range 0.05 M-0.1 M for the CsPbBr$_3$ crystal growth. CsPbBr$_3$ perovskite waveguides were fabricated on glass substrates using a modified soft-lithography approach that combines surface-tension-driven lithography with microfluidics. The perovskite growth process employed an elastomeric PDMS replica obtained by the molding of a silicon master patterned by electron-beam lithography, featuring multiscale micro- and nanostructures with aspect ratios (height/width) ranging from 0.01 to 0.4. The patterns, consisting of arrays of microwire waveguides and interferometer-like geometries, had lateral dimensions ranging from 2 to 30 µm and a height of 250 nm. Each waveguide incorporated arrays of nanometer-scale gratings, with a height of 50 nm, pitches between 250 and 300 nm, and varying filling factors, etched directly onto the surface. A small volume of perovskite precursor solution was injected into the PDMS replica. Crystallization occurred within the microchannels under controlled vapor saturation and temperature. Nucleation initiated at the elastomer–air interface and subsequently progressed via microfluidic-guided growth at controlled temperature. After PDMS removal, CsPbBr$_3$ structures faithfully reproducing the micro- and nanostructures of the master were obtained directly on glass substrates.

### Confocal Microscopy

Confocal images were acquired using a Zeiss LSM 980 confocal microscope in high-resolution mode (1024 x 1024 px). Samples were prepared on microscope slides and characterized using a 488 nm laser and an MBS 488/639 nm plate with appropriate PMT and GaAsP detectors. All samples were imaged at various magnifications using Plan-Apochromat objectives of 10x/0.45, 20x/0.8, and 40x/0.95. Three-dimensional images were obtained using the Z-stack scanning mode, which generates optical slices at various depths. Image analysis across different acquisition channels, tomographic reconstructions, spectral profiles, and 3D rendering were performed using the dedicated ZEN BLUE software.

**Atomic Force Microscopy (AFM) and Scanning Electron Microscopy (SEM)**

AFM measurements were performed with a Dimension Icon (Bruker AXS) instrument, equipped with the Nanoscope V controller. Images were acquired in air, at room temperature, in Tapping Mode to protect the samples from damage. A rotated tapping etched silicon probe (Bruker, Germany) with a nominal radius of curvature $R \leq 8$ nm was used. The morphology of the Si masters was investigated using a Zeiss Sigma 300 (ZEISS) field emission (FE-SEM) scanning electron microscope (FE-SEM) operated at an acceleration voltage of 5 kV and with an in-lens detector.

**Optical Measurements**

All the optical measurements were performed by using a microscope objectives, 100X, N.A. = 0.9 (Zeiss) in a reflection configuration. The pumping lasers are taken from a pulsed source (Carbide, Light Conversion) with a tunable RR and a pulse duration of < 290 fs, which drives two computer-controlled Optical Parametric Amplifier stages (Orpheus OPA, Light Conversion). The OPA outputs are independently tunable in wavelength and, depending on the specific measurement (single-beam or double-beam excitation), have been used either simultaneously or individually. In all the experiments the RR is set to 10 kHz. The dispersions presented in Figure 2, Figure 3 and Supplementary Material are collected by using an off resonance pump (480 nm, 10 kHz) and by imaging the back focal plane of the objective onto the entrance of a spectrometer (Andor, Kymera 328i Spectrograph), equipped with three gratings (150 lines/mm, 300 lines/mm, 600 lines/mm) and coupled to a 2D charge-coupled device (CCD camera, Andor, iStar 334T). The emitted PL signal was filtered using an edgepass filter (Thorlabs, FELH0500) to suppress the residual excitation laser and isolate the PL emission. TE and TM polarizations are defined with the waveguide axis aligned to the entrance slits of the spectrometer, such that the electric field is parallel (TE) or perpendicular (TM) to the grating periodicity. The two polarizations are selectively monitored by using a half-wave plate (AHWP10M-600) and a linear polarizer (GTH10M-A) along the detection line. The laser pumping power is controlled by using a variable neutral-density wheel filter. For double beam experiments, the two OPA outputs are temporally matched by using a delay line (Newport, M-ILS 150CC DC Servo Linear Stage). With reference to Figure 4 c, the temporal profiles are obtained by subtracting the residual PL signal generated by the amplifier alone, which remains constant across all temporal scans. The amplifier-only signal is collected separately using the same exposure time and then subtracted from the measurements in which both beams are simultaneously present.

**Acknowledgments:**

The authors gratefully thank P. Cazzato and P. Aloe for their valuable technical assistance, and M. Barbieri and A. Rastelli for the fruitful discussions. This work is supported by: "Quantum Optical Networks based on Exciton-polaritons", (Q-ONE, N. 101115575, HORIZON-EIC-2022-PATHFINDER CHALLENGES EU project), "National Quantum Science and Technology Institute" (NQSTI, N. PE0000023, PNRR MUR project), "Integrated Infrastructure Initiative in Photonic and Quantum Sciences" (I-PHOQS, N. IR0000016, PNRR MUR project), "Hybrid Perovskite on Silicon CMOS X-ray Detectors" (HyPoSiCX, n. 2022LWHCWY, PRIN2022 MUR project). "Single crystal perovskite nanopixel arrays for miniaturized image sensors" (NanoPix N. 2022YM3232-CUP B53D2300465006 PRIN 2022 MUR Project, funded by the European Community-Next Generation EU, Missione 4, Componente 2).

**Competing interests:**

The authors declare that they have no competing interests.

**Data and materials availability:**

All data needed to evaluate the conclusions are presented in the paper and/or the Supplementary Materials. Additional data related to this paper may be requested from the authors.